# Intuitionistic Neutrosophic Soft Set

Broumi Said[1] and Florentin Smarandache[2]

[1]Administrator of Faculty of Arts and Humanities, Hay El Baraka Ben M'sik Casablanca B.P. 7951, Hassan II University Mohammedia-Casablanca , Morocco
[2]Department of Mathematics, University of New Mexico,
705 Gurley Avenue,Gallup, NM 87301, USA
(*Received December11, 2012, accepted February 24, 2013*)

**Abstract.** In this paper we study the concept of intuitionistic neutrosophic set of Bhowmik and Pal. We have introduced this concept in soft sets and defined intuitionistic neutrosophic soft set. Some definitions and operations have been introduced on intuitionistic neutrosophic soft set. Some properties of this concept have been established.

## 1. Introduction

In wide varities of real problems like , engineering problems, social, economic, computer science, medical science…etc. The data associated are often uncertain or imprecise, all real data are not necessarily crisp, precise, and deterministic because of their fuzzy nature. Most of these problem were solved by different theories, firstly by fuzzy set theory provided by Lotfi , Zadeh [1] ,Later several researches present a number of results using different direction of fuzzy set such as : interval fuzzy set [13], intuitionistic fuzzy set by Atanassov[2], all these are successful to some extent in dealing with the problems arising due to the vagueness present in the real world ,but there are also cases where these theories failed to give satisfactory results, possibly  due to indeterminate and  inconsistent information which exist in belif system, then in 1995, Smarandache [3] intiated the theory of neutrosophic as new mathematical tool for handling problems involving imprecise, indeterminacy,and inconsistent data. Later on  authors like Bhowmik and Pal [7] have further studied the intuitionistic  neutrosophic set and presented various properties of it. In 1999 Molodtsov [4] introduced the concept of soft set which was completely a new approche for dealing with vagueness and uncertainties ,this concept can be seen free from the inadequacy of parameterization tool. After Molodtsovs'work, there have been many researches in combining fuzzy set with soft set, which  incorporates the beneficial properties of both fuzzy set and soft set techniques ( see [12] [6] [8]). Recently , by the concept of neutrosophic set and soft set, first time,  Maji [11] introduced  neutrosophic soft set, established its application in decision making, and thus opened a new direction, new path of thinking to engineers, mathematicians, computer scientists and many others in various tests. This paper is an attempt to combine the concepts: intuitionistic neutrosophic set and soft set together by introducing a new concept called intuitionistic neutrosophic sof set, thus we introduce its operations namely equal ,subset, union ,and intersection, We also   present an application of intuitionistic neutrosophic soft set in decision making problem.

The organization of this paper is as follow : in section 2, we briefly present some basic definitions and preliminary results are given which will be used in the rest of the paper. In section 3, Intuitionistic neutrosophic soft set. In section 4 an application of intuitionistic neutrosophic soft set in a decision making problem. Conclusions are there in the concluding section 5.

## 2. Preliminaries

Throughout this paper, let U be a universal set and E be the set of all possible parameters under consideration with respect to U, usually, parameters are attributes, characteristics, or properties of objects in U. We now recall some basic notions of neutrosophic set, intuitionistic neutrosophic set and soft set.

[1] Corresponding author. Tel.: +2126611416232
 *E-mail address*: broumisaid78@gmail.com

**Definition 2.1 (see[3]).** Let U be an universe of discourse then the neutrosophic set A is an object having the form A = {< x: $T_{A(x)}, I_{A(x)}, F_{A(x)}$ >, x ∈ U}, where the functions T,I,F : U→]$^-0,1^+$[ define respectively the degree of membership, the degree of indeterminacy, and the degree of non-membership of the element x ∈ X to the set A with the condition.

$$^-0 \leq T_{A(x)} + I_{A(x)} + F_{A(x)} \leq 3^+.$$

From philosophical point of view, the neutrosophic set takes the value from real standard or non-standard subsets of ]$^-0,1^+$[.so instead of ]$^-0,1^+$[ we need to take the interval [0,1] for technical applications, because ]$^-0,1^+$[will be difficult to apply in the real applications such as in scientific and engineering problems.

**Definition 2.2 (see [3]).** A neutrosophic set A is contained in another neutrosophic set B i.e. A ⊆ B if ∀x ∈ U, $T_A(x) \leq T_B(x)$, $I_A(x) \leq I_B(x)$, $F_A(x) \geq F_B(x)$.

A complete account of the operations and application of neutrsophic set can be seen in [3] [10].

**Definition 2.3(see[7]). intuitionistic neutrosophic set**

An element x of U is called significant with respect to neutrsophic set A of U if the degree of truth-membership or falsity-membership or indeterminancy-membership value, i.e., $T_{A(x)}$ or $F_{A(x)}$ or $I_{A(x)} \leq 0.5$. Otherwise, we call it insignificant. Also, for neutrosophic set the truth-membership, indeterminacy-membership and falsity-membership all can not be significant. We define an intuitionistic neutrosophic set by A = {< x: $T_{A(x)}, I_{A(x)}, F_{A(x)}$ >, x ∈ U}, where min { $T_{A(x)}, F_{A(x)}$ } ≤ 0.5, min { $T_{A(x)}, I_{A(x)}$ } ≤ 0.5, min { $F_{A(x)}, I_{A(x)}$ } ≤ 0.5, for all x ∈ U, with the condition 0 ≤ $T_{A(x)} + I_{A(x)} + F_{A(x)} \leq 2$.

As an illustration, let us consider the following example.

**Example 2.4.** Assume that the universe of discourse U={$x_1, x_2, x_3$}, where $x_1$ characterizes the capability, $x_2$ characterizes the trustworthiness and $x_3$ indicates the prices of the objects. It may be further assumed that the values of $x_1$, $x_2$ and $x_3$ are in [0,1] and they are obtained from some questionnaires of some experts. The experts may impose their opinion in three components viz. the degree of goodness, the degree of indeterminacy and that of poorness to explain the characteristics of the objects. Suppose A is an intuitionistic neutrosophic set ( IN S ) of U, such that,

A = {< $x_1$,0.3,0.5,0.4 >,< $x_2$,0.4,0.2,0.6 >,< $x_3$,0.7,0.3,0.5 >}, where the degree of goodness of capability is 0.3, degree of indeterminacy of capability is 0.5 and degree of falsity of capability is 0.4 etc.

**Definition 2.5** (see[4]). Let U be an initial universe set and E be a set of parameters. Let P(U) denotes the power set of U. Consider a nonempty set A, A ⊂ E. A pair ( F, A ) is called a soft set over U, where F is a mapping given by F : A → P(U).

As an illustration, let us consider the following example.

**Example 2.6.** Suppose that U is the set of houses under consideration, say U = {$h_1, h_2, \ldots, h_5$}. Let E be the set of some attributes of such houses, say E = {$e_1, e_2, \ldots, e_8$}, where $e_1, e_2, \ldots, e_8$ stand for the attributes "expensive", "beautiful", "wooden", "cheap", "modern", and "in bad repair", respectively.

In this case, to define a soft set means to point out expensive houses, beautiful houses, and so on. For example, the soft set (F, A) that describes the "attractiveness of the houses" in the opinion of a buyer, say Thomas, may be defined like this:

A={$e_1, e_2, e_3, e_4, e_5$};

F($e_1$) = {$h_2, h_3, h_5$}, F($e_2$) = {$h_2, h_4$}, F($e_3$) = {$h_1$}, F($e_4$) = U, F($e_5$) = {$h_3, h_5$}.

For more details on the algebra and operations on intuitionistic neutrosophic set and soft set, the reader may refer to [ 5,6,8,9,12].

## 3. Intuitionistic Neutrosophic Soft Set

In this section, we will initiate the study on hybrid structure involving both intuitionstic neutrosophic set and soft set theory.

**Definition 3.1**. Let U be an initial universe set and A ⊂ E be a set of parameters. Let N( U ) denotes the set of all intuitionistic neutrosophic sets of U. The collection (F,A) is termed to be the soft intuitionistic neutrosophic set over U, where F is a mapping given by F : A → N(U).

**Remark 3.2**. we will denote the intuitionistic neutrosophic soft set defined over an universe by INSS.

Let us consider the following example.

**Example 3.3.** Let U be the set of blouses under consideration and E is the set of parameters (or qualities). Each parameter is a intuitionistic neutrosophic word or sentence involving intuitionistic neutrosophic words. Consider E = { Bright, Cheap, Costly, very costly, Colorful, Cotton, Polystyrene, long sleeve , expensive }. In this case, to define a intuitionistic neutrosophic soft set means to point out Bright blouses, Cheap blouses, Blouses in Cotton and so on. Suppose that, there are five blouses in the universe U given by, U = $\{b_1,b_2,b_3,b_4,b_5\}$ and the set of parameters A = $\{e_1,e_2,e_3,e_4\}$, where each $e_i$ is a specific criterion for blouses:

$e_1$ stands for 'Bright',
$e_2$ stands for 'Cheap',
$e_3$ stands for 'costly',
$e_4$ stands for 'Colorful',

Suppose that,

F(Bright)={<$b_1$,0.5,0.6,0.3>,<$b_2$,0.4,0.7,0.2>,<$b_3$,0.6,0.2,0.3>,<$b_4$,0.7,0.3,0.2>
,<$b_5$,0.8,0.2,0.3>}.

F(Cheap)={<$b_1$,0.6,0.3,0.5>,<$b_2$,0.7,0.4,0.3>,<$b_3$,0.8,0.1,0.2>,<$b_4$,0.7,0.1,0.3>
,<$b_5$,0.8,0.3,0.4>}.

F(Costly)={<$b_1$,0.7,0.4,0.3>,<$b_2$,0.6,0.1,0.2>,<$b_3$,0.7,0.2,0.5>,< $b_4$,0.5,0.2,0.6 >
,< $b_5$,0.7,0.3,0.2 >}.

F(Colorful)={<$b_1$,0.8,0.1,0.4>,<$b_2$,0.4,0.2,0.6>,<$b_3$,0.3,0.6,0.4>,<$b_4$,0.4,0.8,0.5>
,< $b_5$,0.3,0.5,0.7 >}.

The intuitionistic neutrosophic soft set ( INSS ) ( F, E ) is a parameterized family {$F(e_i)$,i = 1,···,10} of all intuitionistic neutrosophic sets of U and describes a collection of approximation of an object. The mapping F here is 'blouses (.)', where dot(.) is to be filled up by a parameter $e_i \in$ E. Therefore, $F(e_1)$ means 'blouses (Bright)' whose functional-value is the intuitionistic neutrosophic set

{< $b_1$,0.5,0.6,0.3 >,< $b_2$,0.4,0.7,0.2 >, < $b_3$,0.6,0.2,0.3 >,< $b_4$,0.7,0.3,0.2 >,< $b_5$,0.8,0.2,0.3 >}.

Thus we can view the intuitionistic neutrosophic soft set ( INSS ) ( F, A ) as a collection of approximation as below:

( F, A ) = { Bright blouses= {< $b_1$,0.5,0.6,0.3 >,< $b_2$,0.4,0.7,0.2 >, < $b_3$,0.6,0.2,0.3 >,< $b_4$,0.7,0.3,0.2 >,< $b_5$,0.8,0.2,0.3 >}, Cheap blouses= {< $b_1$,0.6,0.3,0.5 >,< $b_2$,0.7,0.4,0.3 >,< $b_3$,0.8,0.1,0.2 >, < $b_4$,0.7,0.1,0.3 >,< $b_5$,0.8,0.3,0.4 >}, costly blouses= {< $b_1$,0.7,0.4,0.3 > ,< $b_2$,0.6,0.1,0.2 >,< $b_3$,0.7,0.2,0.5 >,< $b_4$,0.5,0.2,0.6 >,< $b_5$,0.7,0.3,0.2 >}, Colorful blouses= {< $b_1$,0.8,0.1,0.4 >,< $b_2$,0.4,0.2,0.6 >,< $b_3$,0.3,0.6,0.4 >, < $b_4$,0.4,0.8,0.5>,< $b_5$,0.3,0.5,0.7 >}}.

where each approximation has two parts: (i) a predicate p, and (ii) an approximate value-set v ( or simply to be called value-set v ).

For example, for the approximation 'Bright blouses= {< $b_1$,0.5,0.6,0.3 >, < $b_2$,0.4,0.7,0.2 >,<$b_3$,0.6,0.2,0.3>,<$b_4$,0.7,0.3,0.2>,<$b_5$,0.8,0.2,0.3>}'. we have (i) the predicate name 'Bright blouses', and (ii) the approximate value-set is{<$b_1$,0.5,0.6,0.3>,<$b_2$,0.4,0.7,0.2>,<$b_3$,0.6,0.2,0.3>,<$b_4$,0.7,0.3,0.2> ,< $b_5$,0.8,0.2,0.3 >}. Thus, an intuitionistic neutrosophic soft set ( F, E ) can be viewed as a collection of approximation like ( F, E ) = {$p_1 = v_1, p_2 = v_2, \cdots, p_{10} = v_{10}$}. In order to store an intuitionistic neutrosophic soft set in a computer, we could represent it in the form of a table as shown below ( corresponding to the intuitionistic neutrosophic soft set in the above example ). In this table, the entries are $c_{ij}$ corresponding to the blouse $b_i$ and the parameter $e_j$, where $c_{ij}$ = (true-membership value of $b_i$, indeterminacy-membership value of $b_i$, falsity membership value of $b_i$) in $F(e_j)$. The table 1 represent the intuitionistic neutrosophic soft set ( F, A ) described above.

| U | bright | cheap | costly | colorful |
|---|---|---|---|---|
| $b_1$ | ( 0.5,0.6, 0.3 ) | ( 0.6,0.3, 0.5 ) | ( 0.7,0.4, 0.3 ) | ( 0.8,0.1, 0.4 ) |
| $b_2$ | ( 0.4,0.7, 0.2 ) | ( 0.7,0.4, 0.3 ) | ( 0.6,0.1, 0.2 ) | ( 0.4,0.2, 0.6 ) |
| $b_3$ | ( 0.6,0.2, 0.3 ) | ( 0.8,0.1, 0.2 ) | ( 0.7,0.2, 0.5 ) | ( 0.3,0.6, 0.4 ) |
| $b_4$ | ( 0.7,0.3, 0.2 ) | ( 0.7,0.1, 0.3 ) | ( 0.5,0.2, 0.6 ) | ( 0.4,0.8, 0.5 ) |
| $b_5$ | ( 0.8,0.2, 0.3 ) | ( 0.8,0.3, 0.4 ) | ( 0.7,0.3, 0.2 ) | ( 0.3,0.5, 0.7 ) |

Table 1: Tabular form of the INSS ( F, A ).

**Remark 3.4.** An intuitionistic neutrosophic soft set is not an intuituionistic neutrosophic set but a parametrized family of an intuitionistic neutrosophic subsets.

**Definition 3.5. Containment of two intuitionistic neutrosophic soft sets.**
For two intuitionistic neutrosophic soft sets ( F, A ) and ( G, B ) over the common universe U. We say that ( F, A ) is an intuitionistic neutrosophic soft subset of ( G, B ) if and only if
  (i) A ⊂ B.
  (ii) F(e) is an intuitionistic neutrosophic subset of G(e).
     Or $T_{F(e)}(x) \leq T_{G(e)}(x)$, $I_{F(e)}(x) \leq I_{G(e)}(x)$, $F_{F(e)}(x) \geq F_{G(e)}(x)$, $\forall e \in A$, $x \in U$.
We denote this relationship by ( F, A ) ⊆ ( G, B ).
( F, A ) is said to be intuitionistic neutrosophic soft super set of ( G, B ) if ( G, B ) is an intuitionistic neutrosophic soft subset of ( F, A ). We denote it by ( F, A ) ⊇ ( G, B ).

**Example 3.6**. Let (F,A) and (G,B) be two INSSs over the same universe U = $\{o_1,o_2,o_3,o_4,o_5\}$. The INSS (F,A) describes the sizes of the objects whereas the INSS ( G, B ) describes its surface textures. Consider the tabular representation of the INSS ( F, A ) is as follows.

| U | small | large | colorful |
|---|---|---|---|
| $O_1$ | ( 0.4,0.3, 0.6 ) | ( 0.3,0.1, 0.7 ) | ( 0.4,0.1, 0.5 ) |
| $O_2$ | ( 0.3,0.1, 0.4 ) | ( 0.4,0.2, 0.8 ) | ( 0.6,0.3, 0.4 ) |
| $O_3$ | ( 0.6,0.2, 0.5 ) | ( 0.3,0.1, 0.6 ) | ( 0.4,0.3, 0.8 ) |
| $O_4$ | ( 0.5,0.1, 0.6 ) | ( 0.1,0.5, 0.7 ) | ( 0.3,0.3, 0.8 ) |
| $O_5$ | ( 0.3,0.2, 0.4 ) | ( 0.3,0.1, 0.6 ) | ( 0.5,0.2, 0.4 ) |

Table 2: Tabular form of the INSS ( F, A ).

The tabular representation of the INSS ( G, B ) is given by table 3.

| U | small | large | colorful | very smooth |
|---|---|---|---|---|
| O1 | (0.6,0.4, 0.3 ) | ( 0.7,0.2, 0.5 ) | ( 0.5,0.7, 0.4 ) | ( 0.1,0.8, 0.4 ) |
| O2 | ( 0.7,0.5, 0.2 ) | ( 0.4,0.7, 0.3 ) | ( 0.7,0.3, 0.2 ) | ( 0.5,0.7, 0.3 ) |
| O3 | ( 0.6,0.3, 0.5 ) | ( 0.7,0.2, 0.4 ) | ( 0.6,0.4, 0.3 ) | ( 0.2,0.9, 0.4 ) |
| O4 | ( 0.8,0.1, 0.4 ) | ( 0.3,0.6, 0.4 ) | ( 0.4,0.5, 0.7 ) | ( 0.4,0.4, 0.5 ) |
| O5 | ( 0.5,0.4, 0.2 ) | ( 0.4,0.1, 0.5 ) | ( 0.6,0.4, 0.3 ) | ( 0.5,0.8, 0.3 ) |

Table 3: Tabular form of the INSS ( G, B ).

Clearly, by definition 3.5 we have ( F, A ) ⊂ ( G, B ).
**Definition 3.7**. **Equality of two intuitionistic neutrosophic soft sets.**
Two INSSs ( F, A ) and ( G, B ) over the common universe U are said to be intuitionistic neutrosophic soft equal if ( F, A ) is an intuitionistic neutrosophic soft subset of ( G, B ) and ( G, B ) is an intuitionistic neutrosophic soft subset of ( F, A ) which can be denoted by ( F, A )= ( G, B ).
**Definition 3.8**. **NOT set of a set of parameters**.

Let E = {$e_1, e_2, \cdots, e_n$} be a set of parameters. The NOT set of E is denoted by ⏋E is defined by ⏋E ={ ⏋$e_1$, ⏋$e_2$, $\cdots$, ⏋$e_n$}, where ⏋$e_i$ = not $e_i$, $\forall i$ ( it may be noted that ⏋ and ⏋ are different operators ).

**Example 3.9**. Consider the example 3.3. Here ⏋E = { not bright, not cheap, not costly, not colorful }.

**Definition 3.10**. **Complement of an intuitionistic neutrosophic soft set.**

The complement of an intuitionistic neutrosophic soft set ( F, A ) is denoted by $(F,A)^c$ and is defined by $(F,A)^c = (F^c, ⏋A)$, where $F^c : ⏋A \rightarrow N(U)$ is a mapping given by $F^c(\alpha)$ = intuitionistic neutrosophic soft complement with $T^c_{F(x)} = F_{F(x)}, I^c_{F(x)} = I_{F(x)}$ and $F^c_{F(x)} = T_{F(x)}$.

**Example 3.11**. As an illustration consider the example presented in the example 3.2. the complement $(F,A)^c$ describes the 'not attractiveness of the blouses'. Is given below.

F( not bright) = {< $b_1$,0.3,0.6,0.5 >,< $b_2$,0.2,0.7,0.4 >,< $b_3$,0.3,0.2,0.6 >,
            < $b_4$,0.2,0.3,0.7 >< $b_5$,0.3,0.2,0.8 >}.

F( not cheap ) = {< $b_1$,0.5,0.3,0.6 >,< $b_2$,0.3,0.4,0.7 >,< $b_3$,0.2,0.1,0.8 >,
            < $b_4$,0.3,0.1,0.7 >,< $b_5$,0.4,0.3,0.8 >}.

F( not costly ) = {< $b_1$,0.3,0.4,0.7 >,< $b_2$,0.2,0.1,0.6 >,< $b_3$,0.5,0.2,0.7 >,
            < $b_4$,0.6,0.2,0.5 >,< $b_5$,0.2,0.3,0.7 >}.

F( not colorful ) = {< $b_1$,0.4,0.1,0.8 >,< $b_2$,0.6,0.2,0.4 >,< $b_3$,0.4,0.6,0.3 >,
            < $b_4$,0.5,0.8,0.4 >< $b_5$,0.7,0.5,0.3 >}.

**Definition 3.12**: **Empty or Null intuitionistic neutrosopphic soft set.**

An intuitionistic neutrosophic soft set (F,A) over U is said to be empty or null intuitionistic neutrosophic soft (with respect to the set of parameters) denoted by $\Phi_A$ or $(\Phi,A)$ if $T_{F(e)}(m) = 0, F_{F(e)}(m) = 0$ and $I_{F(e)}(m) = 0, \forall m \in U, \forall e \in A$.

**Example 3.13.** Let U = {$b_1, b_2, b_3, b_4, b_5$}, the set of five blouses be considered as the universal set and A = { Bright, Cheap, Colorful } be the set of parameters that characterizes the blouses. Consider the intuitionistic neutrosophic soft set ( F, A) which describes the cost of the blouses and

F(bright)={< $b_1$,0,0,0 >,< $b_2$,0,0,0 >,< $b_3$,0,0,0 >,< $b_4$,0,0,0 >, < $b_5$,0,0,0 >},

F(cheap)={< $b_1$,0,0,0 >,< $b_2$,0,0,0 >,< $b_3$,0,0,0 >,< $b_4$,0,0,0 >, < $b_5$,0,0,0 >},

F(colorful)={< $b_1$,0,0,0 >,< $b_2$,0,0,0 >,< $b_3$,0,0,0 >, < $b_4$,0,0,0 >,< $b_5$,0,0,0 >}.

Here the NINSS ( F, A ) is the null intuitionistic neutrosophic soft set.

**Definition 3.14**. Union of two intuitionistic neutrosophic soft sets.

Let (F,A) and (G,B) be two INSSs over the same universe U. Then the union of (F,A) and (G,B) is denoted by '(F,A)∪(G,B)' and is defined by (F,A)∪(G,B)=(K,C), where C=A∪B and the truth-membership, indeterminacy-membership and falsity-membership of ( K,C) are as follows:

$T_{K(e)}(m)$ = $T_{F(e)}(m)$, if $e \in A - B$,
        = $T_{G(e)}(m)$, if $e \in B - A$,
        = max ($T_{F(e)}(m), T_{G(e)}(m)$), if $e \in A \cap B$.

$I_{K(e)}(m)$ = $I_{F(e)}(m)$, if $e \in A - B$,
        = $I_{G(e)}(m)$, if $e \in B - A$,
        = min ($I_{F(e)}(m), I_{G(e)}(m)$), if $e \in A \cap B$.

$F_{K(e)}(m)$ = $F_{F(e)}(m)$, if $e \in A - B$,
        = $F_{G(e)}(m)$, if $e \in B - A$,
        = min ($F_{F(e)}(m), F_{G(e)}(m)$), if $e \in A \cap B$.

**Example 3.15**. Let ( F, A ) and ( G, B ) be two INSSs over the common universe U. Consider the tabular representation of the INSS ( F, A ) is as follow:

|   | Bright | Cheap | Colorful |
|---|---|---|---|
| $b_1$ | ( 0.6,0.3, 0.5 ) | ( 0.7,0.3, 0.4 ) | ( 0.4,0.2, 0.6 ) |
| $b_2$ | ( 0.5,0.1, 0.8 ) | ( 0.6,0.1, 0.3 ) | ( 0.6,0.4, 0.4 ) |
| $b_3$ | ( 0.7,0.4, 0.3 ) | ( 0.8,0.3, 0.5 ) | ( 0.5,0.7, 0.2 ) |
| $b_4$ | ( 0.8,0.4, 0.1 ) | ( 0.6,0.3, 0.2 ) | ( 0.8,0.2, 0.3 ) |
| b$_5$ | ( 0.6,0.3, 0.2 ) | ( 0.7,0.3, 0.5 ) | ( 0.3,0.6, 0.5 |

Table 4: Tabular form of the INSS ( F, A ).

The tabular representation of the INSS ( G, B ) is as follow:

| U | Costly | Colorful |
|---|---|---|
| b$_1$ | ( 0.6,0.2, 0.3) | ( 0.4,0.6, 0.2 ) |
| b$_2$ | ( 0.2,0.7, 0.2 ) | ( 0.2,0.8, 0.3 ) |
| b$_3$ | ( 0.3,0.6, 0.5 ) | ( 0.6,0.3, 0.4 ) |
| b$_4$ | ( 0.8,0.4, 0.1 ) | ( 0.2,0.8, 0.3 ) |
| b$_5$ | ( 0.7,0.1, 0.4 ) | ( 0.5,0.6, 0.4 ) |

Table 5: Tabular form of the INSS ( G, B ).

Using definition 3.12 the union of two INSS (F, A ) and ( G, B ) is ( K, C ) can be represented into the following Table.

| U | Bright | Cheap | Colorful | Costly |
|---|---|---|---|---|
| b$_1$ | ( 0.6,0.3, 0.5 ) | ( 0.7,0.3, 0.4 ) | ( 0.4,0.2, 0.2 ) | ( 0.6,0.2, 0.3 ) |
| b$_2$ | ( 0.5,0.1, 0.8 ) | ( 0.6,0.1, 0.3 ) | ( 0.6,0.4, 0.3 ) | ( 0.2,0.7, 0.2 ) |
| b$_3$ | ( 0.7,0.4, 0.3 ) | ( 0.8,0.3, 0.5 ) | ( 0.6,0.3, 0.2 ) | ( 0.3,0.6, 0.5 ) |
| b$_4$ | ( 0.8,0.4, 0.1 ) | ( 0.6,0.3, 0.2 ) | ( 0.8,0.2, 0.3 ) | ( 0.8,0.4, 0.1 ) |
| b$_5$ | ( 0.6,0.3, 0.2 ) | ( 0.7,0.3, 0.5 ) | ( 0.5,0.6, 0.4 ) | ( 0.7,0.1, 0.4 ) |

Table 6: Tabular form of the INSS ( K, C ).

**Definition 3.16. Intersection of two intuitionistic neutrosophic soft sets.**
Let (F,A) and (G,B) be two INSSs over the same universe U such that A ∩ B≠0. Then the intersection of (F,A) and ( G,B) is denoted by '( F,A) ∩ (G, B)' and is defined by ( F, A ) ∩( G, B ) = ( K,C),where C =A∩B and the truth-membership, indeterminacy membership and falsity-membership of ( K, C ) are related to those of (F,A) and (G,B) by:

$T_{K(e)}(m)$ = min ($T_{F(e)}(m),T_{G(e)}(m)$),
$I_{K(e)}(m)$ = min ($I_{F(e)}(m),I_{G(e)}(m)$),
$F_{K(e)}(m)$ = max ($F_{F(e)}(m),F_{G(e)}(m)$), for all e ∈ C.

**Example 3.17**. Consider the above example 3.15. The intersection of ( F, A ) and ( G, B ) can be represented into the following table :

| U | Colorful |
|---|---|
| b$_1$ | ( 0.4,0.2,0.6) |
| b$_2$ | ( 0.2,0.4,0.4) |
| b$_3$ | ( 0.6,0.3,0.4) |
| b$_4$ | ( 0.8,0.2,0.3) |
| b$_5$ | ( 0.3,0.6,0.5) |

Table 7: Tabular form of the INSS ( K, C ).

**Proposition 3.18.** If (F, A) and (G, B) are two INSSs over U and on the basis of the operations defined above ,then:

(1) idempotency laws: (F,A) ∪ (F,A) = (F,A).
                     (F,A) ∩ (F,A) = (F,A).
(2) Commutative laws : (F,A) ∪ (G,B) = (G,B) ∪ (F,A).
                      (F,A) ∩ (G,B) = (G,B) ∩ (F,A).
(3) (F,A) ∪ Φ = (F,A).
(4) (F,A) ∩ Φ = Φ.
(5) [(F,A)$^c$]$^c$ = (F,A).

Proof. The proof of the propositions 1 to 5 are obvious.

**Proposition 3.19 .** If ( F, A ), ( G, B ) and ( K, C ) are three INSSs over U,then:
(1) (F,A) ∩ [(G,B) ∩ (K,C)] = [(F,A) ∩ (G,B)] ∩ (K,C).
(2) (F,A) ∪ [(G,B) ∪ (K,C)] = [(F,A) ∪ (G,B)] ∪ (K,C).
(3) Distributive laws: (F,A) ∪ [(G,B) ∩ (K,C)] = [(F,A) ∪ (G,B)] ∩ [(F,A) ∪ (K,C)].
(4) (F,A) ∩ [(G,B) ∪ (K,C)] = [(H,A) ∩ (G,B)] ∪ [(F,A) ∩ (K,C)].

**Exemple 3.20.** Let (F,A) ={⟨$b_1$,0.6,0.3,0. 1⟩,⟨$b_2$,0.4,0.7,0. 5),($b_3$,0.4,0.1,0.8)} , (G,B) ={($b_1$,0.2,0.2,0.6), ($b_2$ 0.7,0.2,0.4), ($b_3$,0.1,0.6,0.7)} and (K,C) ={($b_1$, 0.3,0.8,0.2),⟨$b_2$, 0.4,0.1,0.6),⟨$b_3$,0.9,0.1,0.2)} be three INSSs of U, Then:

(F,A) ∪ (G,B) = { ⟨$b_1$, 0.6,0.2,0.1 ⟩ , ⟨$b_2$, 0.7,0.2,0.4 ⟩ , ⟨$b_3$,0.4,0.1,0.7 ⟩ }.
(F,A) ∪ (K,C) = { ⟨$b_1$,0.6,0.3,0.1 ⟩ , ⟨$b_2$, 0.4,0.1,0.5 ⟩ , ⟨$b_3$,0.9,0.1,0.2 ⟩ }.
(G,B) ∩ (K,C)] = { ⟨$b_1$,0.2,0.2,0.6 ⟩ , ⟨$b_2$,0.4,0.1,0.6 ⟩ , ⟨$b_3$, 0.1,0.1,0.7 ⟩ }.
(F,A) ∪ [(G,B) ∩ (K,C)] = { ⟨$b_1$,0.6,0.2,0.1 ⟩ , ⟨$b_2$,0.4,0.1,0.5 ⟩ , ⟨$b_3$,0.4,0.1,0.7 ⟩ }.
[(F,A) ∪ (G,B)] ∩ [(F,A) ∪ (K,C)] = {⟨$b_1$,0.6,0.2,0.1⟩,⟨$b_2$,0.4,0.1,0.5⟩,⟨$b_3$,0.4,0.1,0.7⟩}.

Hence distributive (3) proposition verified.
Proof, can be easily proved from definition 3.14.and 3.16.

**Definition 3.21**. **AND operation on two intuitionistic neutrosophic soft sets**.
Let ( F, A ) and ( G, B ) be two INSSs over the same universe U. then ( F, A ) ''AND ( G, B) denoted by '( F, A ) ∧ ( G, B )and is defined by ( F, A ) ∧ ( G, B ) = ( K, A × B ), where K(α, β)=F(α)∩ B(β) and the truth-membership, indeterminacy-membership and falsity-membership of ( K, A×B ) are as follows:

$$T_{K(α,β)}(m) = \min(T_{F(α)}(m),T_{G(β)}(m)), I_{K(α,β)}(m) = \min(I_{F(α)}(m),I_{G(β)}(m))$$
$$F_{K(α,β)}(m) = \max(F_{F(α)}(m),F_{G(β)}(m)), ∀α∈ A, ∀β∈ B.$$

**Example 3.22**. Consider the same example 3.15 above. Then the tabular representation of (F,A) AND ( G, B ) is as follow:

| u | (bright, costly) | (bright, Colorful) | (cheap, costly) |
|---|---|---|---|
| $b_1$ | ( 0.6,0.2, 0.5 ) | ( 0.4,0.3, 0.5 ) | ( 0.6,0.2, 0.4 ) |
| $b_2$ | ( 0.2,0.1, 0.8 ) | ( 0.2,0.1, 0.8 ) | ( 0.2,0.1, 0.3 ) |
| $b_3$ | ( 0.3,0.4, 0.5 ) | ( 0.6,0.3, 0.4 ) | ( 0.3,0.3, 0.5 ) |
| $b_4$ | ( 0.8,0.4, 0.1 ) | ( 0.2,0.4,0.3 ) | ( 0.6,0.3, 0.2 ) |
| $b_5$ | ( 0.6,0.1, 0.4 ) | ( 0.5,0.3, 0.4 ) | ( 0.7,0.1, 0.5) |
| u | (cheap, colorful) | (colorful, costly) | (colorful, colorful) |
| $b_1$ | ( 0.4,0.3, 0.4 ) | ( 0.4,0.2, 0.6 ) | ( 0.4,0.2, 0.6 ) |
| $b_2$ | ( 0.2,0.1, 0.3 ) | ( 0.2,0.4, 0.4 ) | ( 0.2,0.4, 0.4 ) |
| $b_3$ | ( 0.6,0.3, 0.5 ) | ( 0.3,0.6, 0.5 ) | ( 0.5,0.3, 0.4 ) |
| $b_4$ | ( 0.2,0.3, 0.3 ) | ( 0.8,0.2, 0.3 ) | ( 0.2,0.2, 0.3 ) |
| $b_5$ | ( 0.5,0.3, 0.5 ) | ( 0.3,0.1, 0.5 ) | ( 0.3,0.6, 0.5 ) |

Table 8: Tabular representation of the INSS ( K, A × B).

**Definition 3.23**. If (F,A) and (G,B) be two INSSs over the common universe U then '(F,A) OR(G,B)' denoted by (F,A) ∨ (G,B) is defined by ( F, A) ∨ (G, B ) = (O,A×B), where, the truth-membership, indeterminacy membership and falsity-membership of O( α, β) are given as follows:

$TO(\alpha,\beta)^{(m)} = \max(T_{F(\alpha)}^{(m)}, T_{G(\beta)}^{(m)})$,
$I_{O(\alpha,\beta)}(m) = \min(I_{F(\alpha)}^{(m)}, I_{G(\beta)}^{(m)})$,
$FO(\alpha,\beta)^{(m)} = \min(F_{F(\alpha)}^{(m)}, F_{G(\beta)}^{(m)}), \forall \alpha \in A, \forall \beta \in B$.

**Example 3.24** Consider the same example 3.14 above. Then the tabular representation of ( F, A ) OR ( G, B ) is as follow:

| u | (bright, costly) | (bright, colorful) | (cheap, costly) |
|---|---|---|---|
| $b_1$ | ( 0.6, 0.2, 0.3 ) | ( 0.6, 0.3, 0.2 ) | ( 0.7, 0.2, 0.3 ) |
| $b_2$ | ( 0.5, 0.1, 0.2 ) | ( 0.5, 0.1, 0.3 ) | ( 0.6, 0.1, 0.2 ) |
| $b_3$ | ( 0.7, 0.4, 0.3 ) | ( 0.7, 0.3, 0.3 ) | ( 0.8, 0.3, 0.5 ) |
| $b_4$ | ( 0.8, 0.4, 0.1 ) | ( 0.8, 0.4, 0.1 ) | ( 0.8, 0.3, 0.1 ) |
| $b_5$ | ( 0.7, 0.1, 0.2 ) | ( 0.6, 0.3, 0.4 ) | ( 0.7, 0.1, 0.4 ) |
| u | (cheap, colorful) | (colorful, costly) | (colorful, colorful) |
| $b_1$ | ( 0.7, 0.3, 0.2 ) | ( 0.6, 0.2, 0.3 ) | ( 0.4, 0.2, 0.2 ) |
| $b_2$ | ( 0.6, 0.1, 0.3 ) | ( 0.6, 0.4, 0.2 ) | ( 0.6, 0.4, 0.3 ) |
| $b_3$ | ( 0.8, 0.3, 0.4 ) | ( 0.5, 0.6, 0.2 ) | ( 0.5, 0.7, 0.2 ) |
| $b_4$ | ( 0.6, 0.3, 0.2 ) | ( 0.8, 0.2, 0.1 ) | ( 0.8, 0.2, 0.3 ) |
| $b_5$ | ( 0.7, 0.3, 0.4 ) | ( 0.7, 0.1, 0.4 ) | ( 0.5, 0.6, 0.4) |

Table 9: Tabular representation of the INSS ( O, A × B).

**Proposition 3.25.** if ( F, A ) and ( G, B ) are two INSSs over U, then :

(1) $[(F,A) \wedge (G,B)]^c = (F,A)^c \vee (G,B)^c$

(2) $[(F,A) \vee (G,B)]^c = (F,A)^c \wedge (G,B)^c$

**Proof 1.** Let $(F,A) = \{<b, T_{F(x)}(b), I_{F(x)}(b), F_{F(x)}(b)> | b \in U\}$
and
$$(G,B) = \{<b, T_{G(x)}(b), I_{G(x)}(b), F_{G(x)}(b)> | b \in U\}$$
be two INSSs over the common universe U. Also let $(K, A \times B) = (F,A) \wedge (G,B)$,
where, $K(\alpha, \beta) = F(\alpha) \cap G(\beta)$ for all $(\alpha, \beta) \in A \times B$ then
$K(\alpha, \beta) = \{<b, \min(T_{F(\alpha)}(b), T_{G(\beta)}(b)), \min(I_{F(\alpha)}(b), I_{G(\beta)}(b)), \max(F_{F(\alpha)}(b), F_{G(\beta)}(b)) > | b \in U\}$.
Therefore,
$[(F,A) \wedge (G,B)]^c = (K, A \times B)^c$
$= \{<b, \max(F_{F(\alpha)}(b), F_{G(\beta)}(b)), \min(I_{F(\alpha)}(b), I_{G(\beta)}(b)), \min(T_{F(\alpha)}(b), T_{G(\beta)}(b)) > | b \in U\}$.
Again
$(F,A)^c \vee (G,B)^c$
$= \{<b, \max(F_{F(\alpha)}^c(b), F_{G(\beta)}^c(b)), \min(I_{F(\alpha)}^c(b), I_{G(\beta)}^c(b)), \min(T_{F(\alpha)}^c(b), T_{G(\beta)}^c(b)) > | b \in U\}$.
$= \{<b, \min(T_{F(\alpha)}(b), T_{G(\beta)}(b)), \min(I_{F(\alpha)}(b), I_{G(\beta)}(b)), \max(F_{F(\alpha)}(b), F_{G(\beta)}(b)) > | b \in U\}^c$
$= \{<b, \max(F_{F(\alpha)}(b), F_{G(\beta)}(b)), \min(I_{F(\alpha)}(b), I_{G(\beta)}(b)), \min(T_{F(\alpha)}(b), T_{G(\beta)}(b)) > | b \in U\}$.

It follows that $[(F,A) \wedge (G,B)]^c = (F,A)^c \vee (G,B)^c$.

**Proof 2.**
Let $(F, A) = \{<b, T_{F(x)}(b), I_{F(x)}(b), F_{F(x)}(b) > | b \in U\}$ and
$(G,B) = \{<b, T_{G(x)}(b), I_{G(x)}(b), F_{G(x)}(b) > | b \in U\}$ be two INSSs over the common universe U. Also let $(O, A \times B) = (F,A) \vee (G,B)$, where, $O(\alpha, \beta) = F(\alpha) \cup G(\beta)$ for all $(\alpha, \beta) \in A \times B$. then
$O(\alpha, \beta) = \{<b, \max(T_{F(\alpha)}(b), T_{G(\beta)}(b)), \min(I_{F(\alpha)}(b), I_{G(\beta)}(b)), \min(F_{F(\alpha)}(b), F_{G(\beta)}(b)) > | b \in U\}$.
$[(F,A) \vee (G,B)]^c = (O, A \times B)^c = \{<b, \min(F_{F(\alpha)}(b), F_{G(\beta)}(b)), \min(I_{F(\alpha)}(b), I_{G(\beta)}(b)),$
$\max(T_{F(\alpha)}(b), T_{G(\beta)}(b)) > | b \in U\}$.
Again
$(H,A)^c \wedge (G,B)^c$
$= \{<b, \min(F_{F(\alpha)}^c(b), F_{G(\beta)}^c(b)), \min(I_{F(\alpha)}^c(b), I_{G(\beta)}^c(b)), \max(T_{F(\alpha)}^c(b), T_{G(\beta)}^c(b)), > | b \in U\}$.

= {< b,max($T_{F(\alpha)}$(b),$T_{G(\beta)}$(b)),min($I_F^c{}_{(\alpha)}$(b),$I_G^c{}_{(\beta)}$(b)),min($F_{F(\alpha)}$(b),$F_{G(\beta)}$(b))>| b ∈ U}$^c$ .

= {< b, min($F_{F(\alpha)}$(b),$F_{G(\beta)}$(b)),min($I_{F(\alpha)}$(b),$I_{G(\beta)}$(b)), max($T_{F(\alpha)}$(b),$T_{G(\beta)}$(b)) >| b ∈ U}.

It follows that [(F,A) ∨ (G,B)]$^c$ = (F,A)$^c$ ∧ (G,B)$^c$ .

## 4. An application of intuitionistic neutrosophic soft set in a decision making problem

For a concrete example of the concept described above, we revisit the blouse purchase problem in Example 3.3. So let us consider the intuitionistic neutrosophic soft set S = (F,P) (see also Table 10 for its tabular representation), which describes the "attractiveness of the blouses" that Mrs. X is going to buy. on the basis of her m number of parameters ($e_1,e_2,…,e_m$) out of n number of blouses($b_1,b_2,…,b_n$). We also assume that corresponding to the parameter $e_j$(j =1,2,···,m) the performance value of the blouse $b_i$ (i = 1,2,···,n) is a tuple $p_{ij}$ = ($T_{F(ej)}$ ($b_i$),$I_{F(ej)}$ ($b_i$),$T_{F(ej)}$ ($b_i$)), such that for a fixed i that values $p_{ij}$ (j = 1,2,···,m) represents an intuitionistic neutrosophic soft set of all the n objects. Thus the performance values could be arranged in the form of a matrix called the 'criteria matrix'. The more are the criteria values, the more preferability of the corresponding object is. Our problem is to select the most suitable object i.e. the object which dominates each of the objects of the spectrum of the parameters $e_j$. Since the data are not crisp but intuitionistic neutrosophic soft the selection is not straightforward. Our aim is to find out the most suitable blouse with the choice parameters for Mrs. X. The blouse which is suitable for Mrs. X need not be suitable for Mrs. Y or Mrs. Z, as the selection is dependent on the choice parameters of each buyer. We use the technique to calculate the score for the objects.

### 4.1. Definition: Comparison matrix

The Comparison matrix is a matrix whose rows are labelled by the object names of the universe such as $b_1,b_2,···,b_n$ and the columns are labelled by the parameters $e_1,e_2,···,e_m$. The entries are $c_{ij}$, where $c_{ij}$, is the number of parameters for which the value of $b_i$ exceeds or is equal to the value $b_j$. The entries are calculated by $c_{ij}$ =a + d - c, where 'a' is the integer calculated as 'how many times $T_{bi}$ ($e_j$) exceeds or equal to $T_{bk}$ ($e_j$)', for $b_i ≠ b_k$, ∀ $b_k$ ∈ U, 'd' is the integer calculated as 'how many times $I_{bi(ej)}$ exceeds or equal to $I_{bk(ej)}$', for $b_i ≠ b_k$, ∀ $b_k$ ∈ U and 'c' is the integer 'how many times $F_{bi(ej)}$ exceeds or equal to $F_{bk}(e_j)$', for $b_i ≠ b_k$, ∀ $b_k$ ∈ U.

**Definition 4.2.** Score of an object. The score of an object $b_i$ is $S_i$ and is calculated as :

$$S_i = \sum_j c_{ij}$$

Now the algorithm for most appropriate selection of an object will be as follows.

Algorithm

(1) input the intuitionistic Neutrosophic Soft Set ( F, A).
(2) input P, the choice parameters of Mrs. X which is a subset of A.
(3) consider the INSS ( F, P) and write it in tabular form.
(4) compute the comparison matrix of the INSS ( F, P).
(5) compute the score $S_i$ of $b_i$,∀i.
(6) find $S_k$ = maxi $S_i$
(7) if k has more than one value then any one of $b_i$ may be chosen.

To illustrate the basic idea of the algorithm, now we apply it to the intuitionistic neutrosophic soft set based decision making problem.

Suppose the wishing parameters for Mrs. X where P={Bright,Costly, Polystyreneing,Colorful,Cheap}.

Consider the INSS ( F, P ) presented into the following table.

| U | Bright | costly | Polystyreneing | Colorful | Cheap |
|---|---|---|---|---|---|
| $b_1$ | ( 0.6,0.3, 0.4 ) | ( 0.5,0.2, 0.6 ) | ( 0.5,0.3, 0.4 ) | ( 0.8,0.2, 0.3 ) | ( 0.6,0.3, 0.2 ) |
| $b_2$ | ( 0.7,0.2, 0.5 ) | ( 0.6,0.3, 0.4 ) | ( 0.4,0.2, 0.6 ) | ( 0.4,0.8, 0.3 ) | ( 0.8,0.1, 0.2 ) |
| $b_3$ | ( 0.8,0.3, 0.4 ) | ( 0.8,0.5, 0.1 ) | ( 0.3,0.5, 0.6 ) | ( 0.7,0.2, 0.1 ) | ( 0.7,0.2, 0.5 ) |
| $b_4$ | ( 0.7,0.5, 0.2 ) | ( 0.4,0.8, 0.3 ) | ( 0.8,0.2, 0.4 ) | ( 0.8,0.3, 0.4 ) | ( 0.8,0.3, 0.4 ) |
| $b_5$ | ( 0.3,0.8, 0.4 ) | ( 0.3,0.6, 0.1 ) | ( 0.7,0.3, 0.2 ) | ( 0.6,0.2, 0.4 ) | ( 0.6,0.4, 0,2 ) |

Table 10: Tabular form of the INSS (F, P).

The comparison-matrix of the above INSS ( F, P) is represented into the following table.

| U | Bright | Costly | Polystyreneing | Colorful | Cheap |
|---|---|---|---|---|---|
| $b_1$ | 0 | -2 | 3 | 0 | 2 |
| $b_2$ | -1 | 1 | -2 | 2 | 2 |
| $b_3$ | 3 | 5 | 0 | 4 | -1 |
| $b_4$ | 6 | 3 | 3 | 3 | 4 |
| $b_5$ | 7 | 2 | 6 | -1 | 3 |

Table 11: Comparison matrix of the INSS ( F, P ).

Next we compute the score for each $b_i$ as shown below:

| U | Score ($S_i$) |
|---|---|
| $b_1$ | 3 |
| $b_2$ | 2 |
| $b_3$ | 11 |
| $b_4$ | 19 |
| $b_5$ | 17 |

Clearly, the maximum score is the score 19, shown in the table above for the blouse $b_4$.
Hence the best decision for Mrs. X is to select $b_4$, followed by $b_5$.

## 5. Conclusions

In this paper we study the notion of intuitionistic neutrosophic set initiated by Bhowmik and Pal. We use this concept in soft sets considering the fact that the parameters ( which are words or sentences ) are mostly intuitionistic neutrosophic set; but both the concepts deal with imprecision, We have also defined some operations on INSS and prove some propositions. Finally, we present an application of INSS in a decision making problem.

## Acknowledgements.

The authors are thankful to the anonymous referee for his valuable and constructive remarks that helped to improve the clarity and the completeness of this paper.